\documentclass[twoside]{ilcws08proc_latex/ilcws08}
\usepackage[latin1]{inputenc}
\usepackage[dvips]{graphicx,epsfig,color}
\usepackage{wrapfig,rotating}
\usepackage{amssymb,amsmath,array}
\usepackage{pstricks}

\begin{document}
\title{
%%%%   Paper title goes here  %%%%%%%%%%%%%%
ILC Main Linac Alignment Simulations using Conventional Techniques and the Rapid Tunnel Reference Survey Model (RTRSM)} %% 
%***********************************************************************
% AUTHORS INFORMATION AREA
%***********************************************************************
\author{\underline{John Dale}, Armin Reichold
% Optional short acknowledgment: remove next line if non-needed
\\\\
%\vspace{.3cm}\\
% Addresses and institutions (remove "1- " in case of a single institution)
John Adams Institute for Accelerator Physics,
University of Oxford, UK
%% Remove the next three lines in case of a single institution
%\vspace{.1cm}\\
\\
%2- Second Author's Institution - Department \\
%Address of Second Author's Institution - Country\\
}
%%***********************************************************************
% END OF AUTHORS INFORMATION AREA
%***********************************************************************

\maketitle

\begin{abstract}
Alignment of the ILC main linac will be more critical than for any currently existing accelerators due to its long length and the ultra low emittance required.  There are several techniques for measuring the ILC reference network; in this report conventional methods for measuring the network and aligning the main linac are simulated. Dispersion Matched Steering (DMS) is applied to the simulated accelerators to determine their final emittance.  Simulations and reconstructions of the ILC reference network usually require very resource intensive programs, the RTRSM has been developed to rapidly generate networks with the required statistical properties.  The RTRSM is discussed in this report and some problems are demonstrated.

\end{abstract}

\section{Introduction}

To achieve the high luminosity required by the ILC, the beams need to have ultra low emittance at the interaction point.  The primary causes of emittance growth in the main linac are component misalignment errors. In sections \ref{conventionalTechnique} to \ref{simulations} the effect of component misalignment resulting from conventional alignment techniques are discussed.

To design and study the ILC a large number of beam dynamics simulations are required, many of which will need to take account of misalignments. Consequently numerous reference networks need to be simulated. Full simulations and reconstructions of realistic ILC reference networks are very resource intensive and therefore the RTRSM has been developed to simulate reference networks rapidly and with the same statistical properties as those produced by full simulations $^{\cite{Freddy_simul}}$. A study of the RTRSM is presented in section \ref{RTRSM}.

\section{A Conventional Technique for the Alignment of the ILC}
\label{conventionalTechnique}

There are many possible techniques for the measurement of the ILC reference network. The conventional reference network measurement method studied in this report uses a laser tracker to measure wall markers (see section \ref{trackers}), a small number of which are measured again using GPS (see section \ref{gps}).  All of the measurements are described by a linear algebra model and processed by a solver (adjusted) to determine the final marker positions (see section \ref{adjust}).  The adjusted network is then used to misalign the simulated ILC (see section \ref{align}) for beam dynamics studies.

\subsection{Laser tracker measurements}
\label{trackers}
The laser tracker is approximately placed at a predetermined position in the tunnel and records the position of a set of Reference Markers (RMs). It is then moved a short distance to measure a different set of RMs which overlaps with the previous section. This is repeated down the entire length of the main linac tunnel (see section \ref{layout}).  

\subsection{GPS measurements}
\label{gps}
On the surface above the tunnel we assume there to be a network of GPS measurement stations which can determine their position with respect to each other. The GPS position information is then transfered down into the tunnel via access shafts. The transfered GPS is then used to determines the position of a RM at the bottom one access shaft with respect to the RMs at the bottom of other access shafts. RMs measured by GPS are referred to as Primary Reference Markers (PRMs). The transfer process through the shafts is not simulated here in detail, instead an error for the combined GPS measurement and transfer is assumed (see table \ref{tab:SimulationParameters}).

\subsection{Network Adjustment}
\label{adjust}
The final RM positions can be determined by a process called network adjustment. Network adjustment takes all of the measurements in the network (for example GPS and Laser tracker measurements) and uses a linear algebra model to determine the best fit RM positions.  In this report network adjustment is performed by PANDA \cite{Panda}. PANDA is a software package which can design, optimize, adjust (solve for positions) and assess 3D networks. It is a commercial package used by, for example, the DESY geodesy group \cite{DESY_Geodesy}.  PANDA produces the position and position error of all the RMs.

\subsection{Main Linac Alignment against the Reference Network}
\label{align}
With the reference network determined, the components of the main linac can be aligned.  In practice this is done by placing a laser tracker in the tunnel near the component to be aligned.  The laser tracker measures its position with respect to the local RMs, and using the knowledge of the adjusted network, determines its own position.  It then uses this information to determine the components position which is then moved to the required location.

For the simulations covered in this report a simplified method is used; this method simply uses the 3 closest RM rings (see section \ref{layout}) and fits a straight line to them.  The fitted straight line is then used to determine the position of the component.

\section{Machine Simulation and Dispersion Matched Steering (DMS)}

The positron side of the ILC main linac is simulated using the Merlin C++ library \cite{Merlin} based package ILCDFS \cite{Merlin_ilc}. Simulated components are positioned within cryomodules with random errors given in table \ref{tab:compErrors}. The supports of the cryomodules are aligned with respect to the reference network. The machine is also set to follow the Earths curvature and as a consequence some dispersion is expected which means that dispersion should be matched to its design value; not eliminated.

The DMS was set up with the following parameters : initial beam emittance of 20nm, nominal beam starting energy of 15GeV accelerating to 250GeV, the initial energy of the test beam of -20\% and constant gradient adjustment of -20\%. This has been shown to be effective in obtaining a low emittance \cite{Merlin_DFS}.

\begin{table}
	\centering
		\begin{tabular}{| c | c |}
		\hline
		Components & Error Values (RMS) \\
		\hline
		Cavity Offset & 300$\mu$\\
		Cavity Tilt & 300$\mu$\\
		Quadrupole Offset & 300$\mu$\\
		Quadrupole roll & 300$\mu$\\
		BPM Offset & 200$\mu$ \\
		\hline	
		\end{tabular}
	\caption{Component static gaussian errors with respect to the module axis}
	\label{tab:compErrors}
\end{table}

\section{Reference Network Layout}
\label{layout}

\begin{figure}
\begin{minipage}[b]{0.5\linewidth}
	\centering
		\includegraphics[width=\columnwidth]{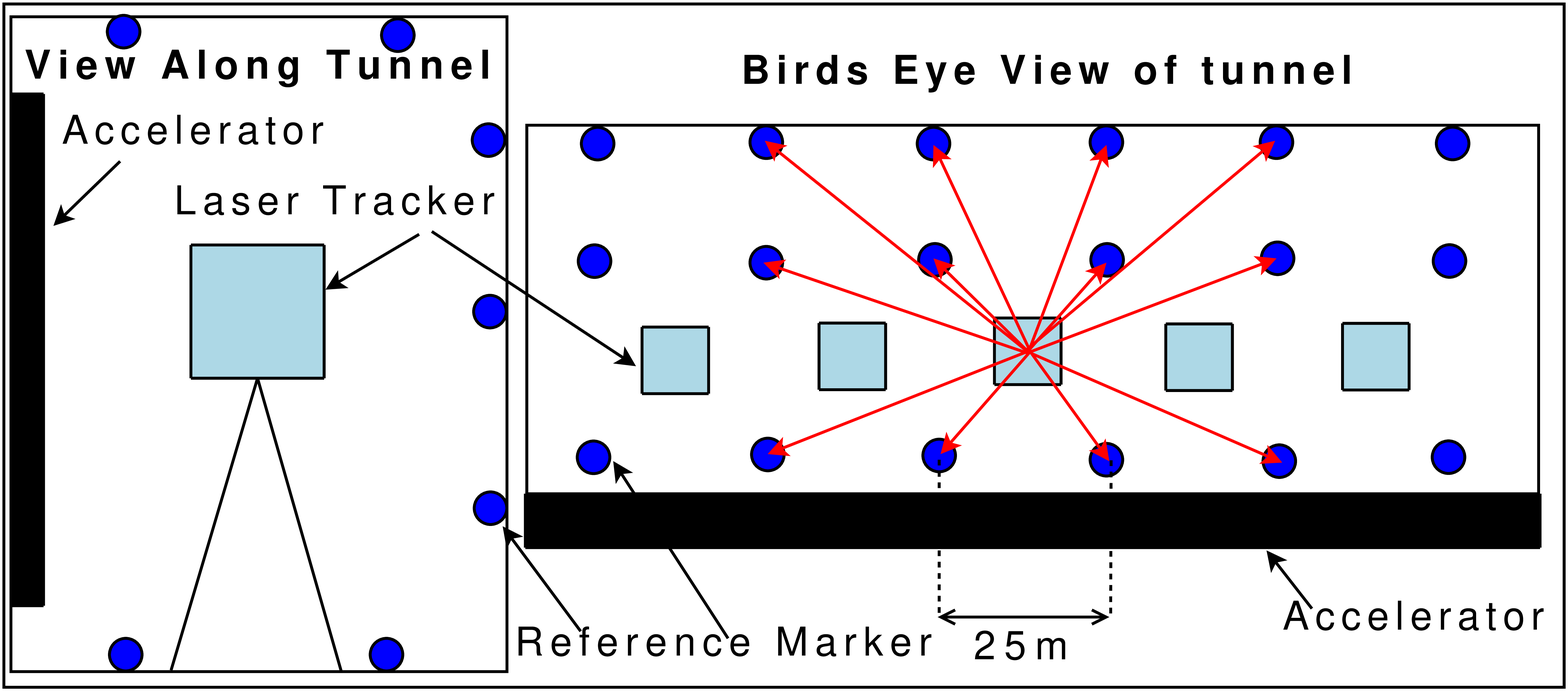}
	\caption{The Reference Marker configuration shown looking down the tunnel and looking along the tunnel}
	\label{markerRing}
	\end{minipage}
	\hspace{0.5cm} % To get a little bit of space between the figures
	\begin{minipage}[b]{0.5\linewidth}
	\centering
		\includegraphics[width=\columnwidth]{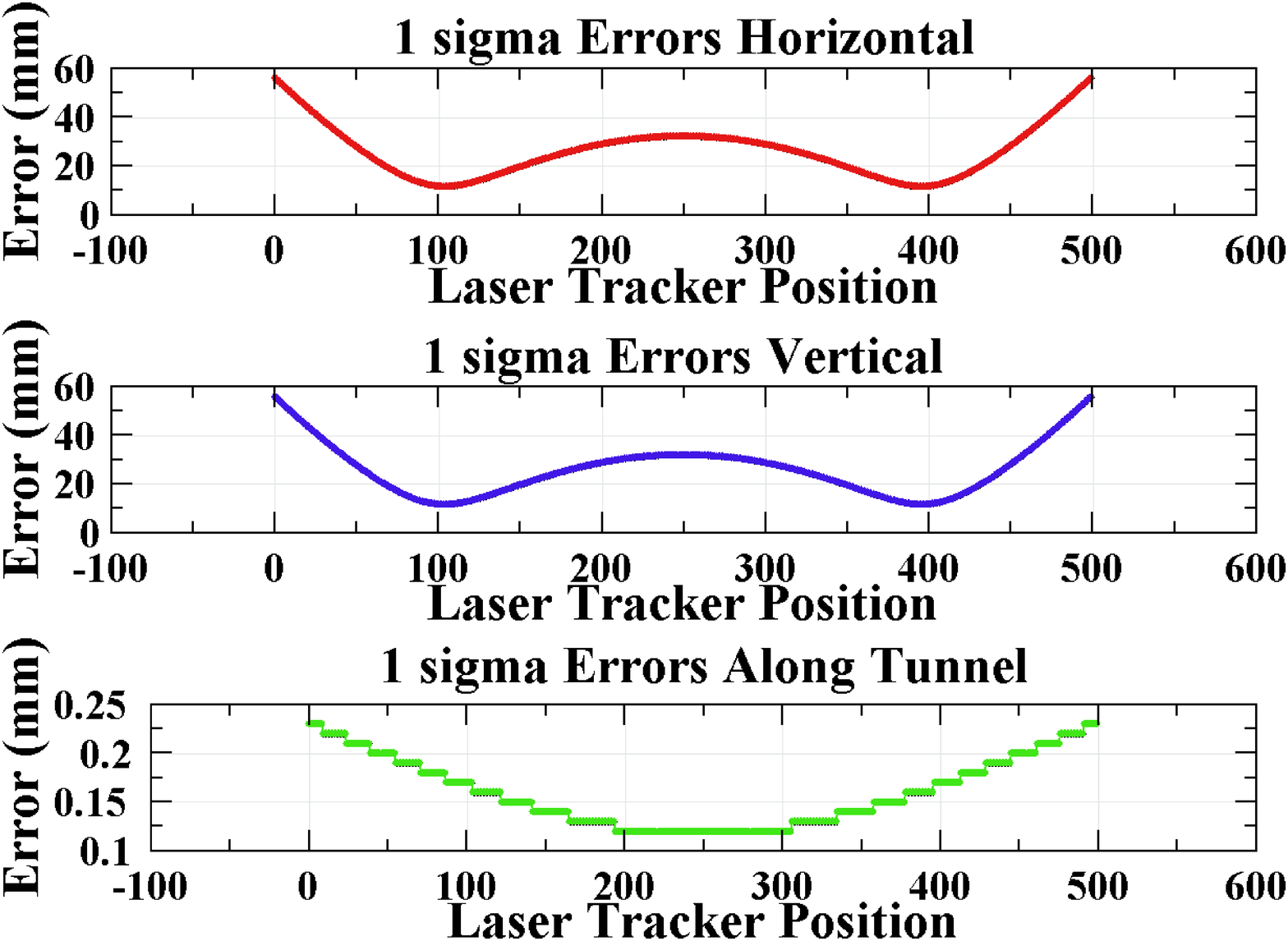}%
	\caption{The Errors on the reference network without PRM}
	\label{NetorkErrors_NoPRM}
	\end{minipage}
\end{figure}

The reference network used in the simulations is made up of rings of RMs.  A ring consists of 7 RMs: 2 on the ceiling, 2 on the floor and 3 on the wall away from the accelerator (see figure~\ref{markerRing}). It is assumed that the accelerator will block the view to any RMs on the wall behind it and for simplicity the tunnel is assumed to be rectangular; which is not expected to affect the results significantly. A ring is placed every 25m along the tunnel, a laser tracker is placed between each ring and measures the two nearest rings in both directions (see figure~\ref{markerRing}). 

\section{Misalignment Simulations}
\label{simulations}

Misalignment simulations were preformed in two groups, one without PRMs and another with PRMs.  The simulations consisted of generating all of the laser tracker measurements and/or PRM measurements using a JAVA program.  The JAVA program did not apply any systematic errors; this will make the measurements and therefore the adjustments optimistic as, for example, refraction in air can be very significant. The simulated measurements were smeared by the relevent resolutions (see table \ref{tab:SimulationParameters} for parameters) and PANDA was used to determine the final RM positions. A ILC main linac is then simulated and misaligned in the vertical plane using the adjusted network. Only the vertical emittance is studied as this is the most critical factor to the ILC performance.

%The vertical adjusted network was used to misalign the simulated ILC main linac because only the vertical emittance is studied as this is the most critical factor to the ILC performance.

\begin{table}
	\centering
		\begin{tabular}{|c |c |c |c|}
		\hline
		\multicolumn{3}{|c|}{Laser Tracker Parameters} & PRM parameters\\
		\hline
			Distance & Azimuth & Zenith & PRM uncorrelated errors\\
			 & & & $\Delta$x, $\Delta$y, $\Delta$z\\
			\hline
			0.1mm+0.5ppm & 0.3 mgon (4.7 $\mu$rad) & 0.3 mgon (4.7 $\mu$rad) & 10mm\\
		\hline	
		\end{tabular}
	\caption{Simulation Parameters : Note PRM errors are not correlated}
	\label{tab:SimulationParameters}
\end{table}

\subsection{Simulations without PRM}

The simulations were initially performed without PRMs. With the results presented in sections \ref{adjustNoPRM} and \ref{dmsNoGPS}.

\subsubsection{Network Adjustment Results}
\label{adjustNoPRM}
PANDA was initially used to determine the errors on the adjusted network (see figure \ref{NetorkErrors_NoPRM}).  The error curves shown in figure \ref{NetorkErrors_NoPRM} follow a double minima curve as expected.

\begin{figure}
	\begin{minipage}[b]{0.5\linewidth} % A minipage that covers half the page
		\centering
		\includegraphics[width=\columnwidth]{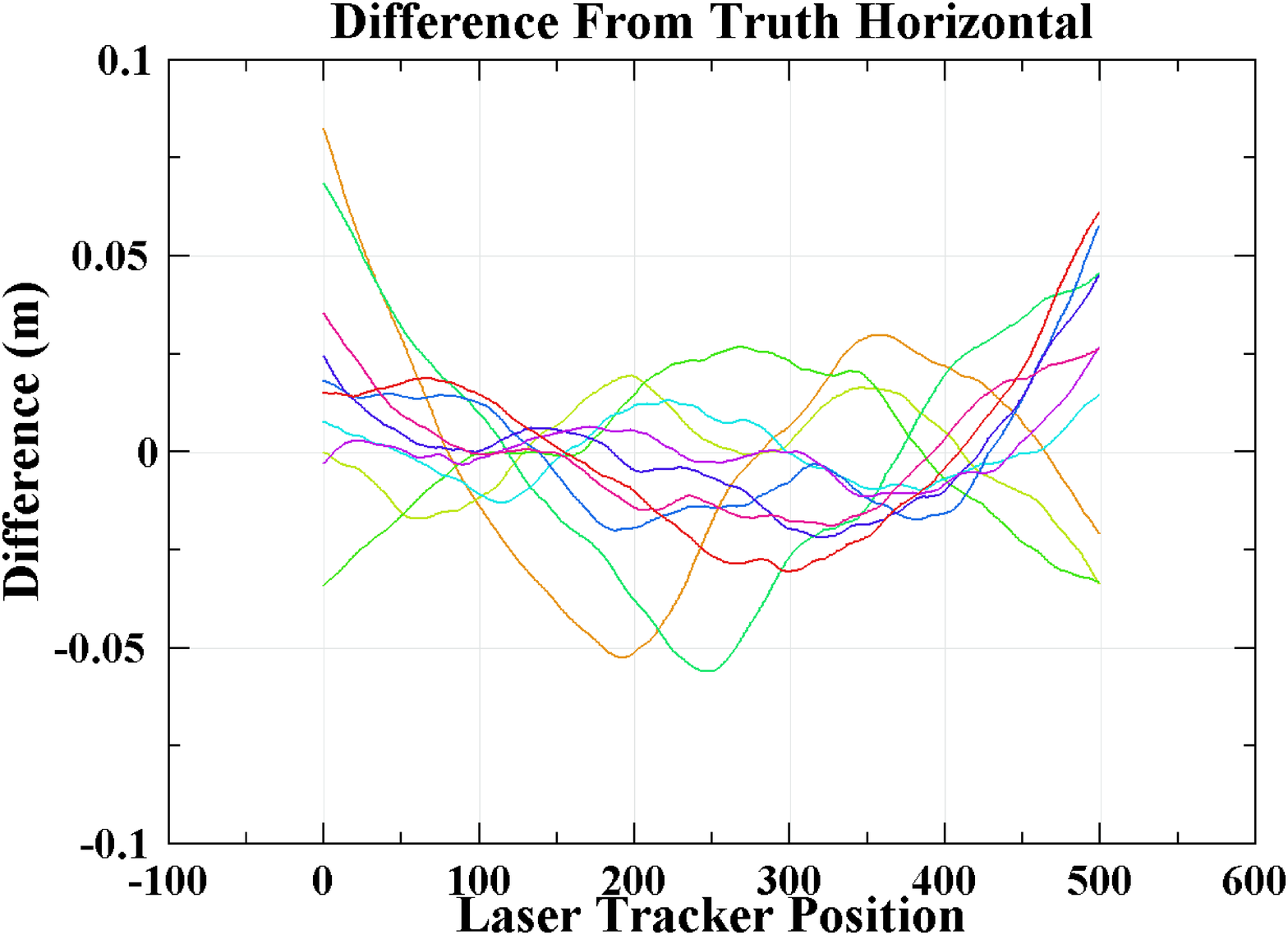}
		\caption{The Horizontal Difference from truth without PRM}
		\label{Horizontal_NoPRM}
	\end{minipage}
\hspace{0.5cm} % To get a little bit of space between the figures
	\begin{minipage}[b]{0.5\linewidth}
		\centering
		\includegraphics[width=\columnwidth]{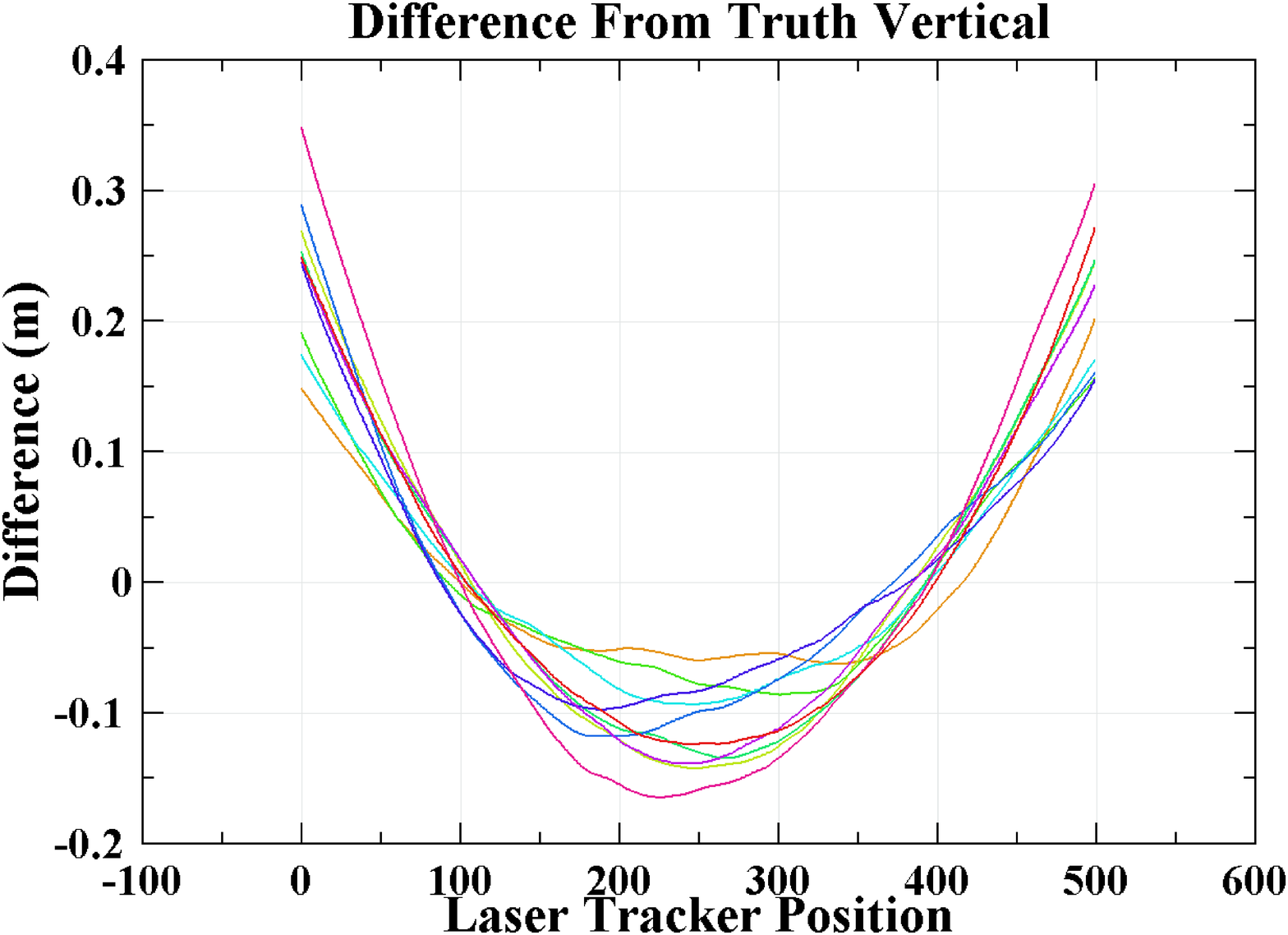}
		\caption{The Vertical Difference from truth without PRM}
		\label{Vertical_NoPRM}
	\end{minipage}
\end{figure}

Ten reference networks were adjusted; the difference from the truth in the horizontal and vertical plane for each run is shown in figures \ref{Horizontal_NoPRM} and \ref{Vertical_NoPRM} respectively. Figure \ref{Horizontal_NoPRM} appears as expected, given the horizontal errors shown in figure \ref{NetorkErrors_NoPRM}.  Figure \ref{Vertical_NoPRM}, does not appear as expected, given the vertical errors shown in figure \ref{NetorkErrors_NoPRM}, it has a large systematic curvature and clearly indicates a problem with the PANDA network adjustment (Note: this problem has since been addressed by the authors of PANDA). The systematic curvature could lead to problems as we wish to study the vertical emittance using DMS; however, if we compare the horizontal and vertical errors from figure \ref{NetorkErrors_NoPRM}, we see that they have the same shape and size.  This allows us to use the horizontal network positions as the vertical network positions in the DMS simulations and work around the PANDA problem.

\subsubsection{DMS Results}
\label{dmsNoGPS}

\begin{figure}
\begin{minipage}[b]{0.5\linewidth}
	\centering
		\includegraphics[width=\columnwidth]{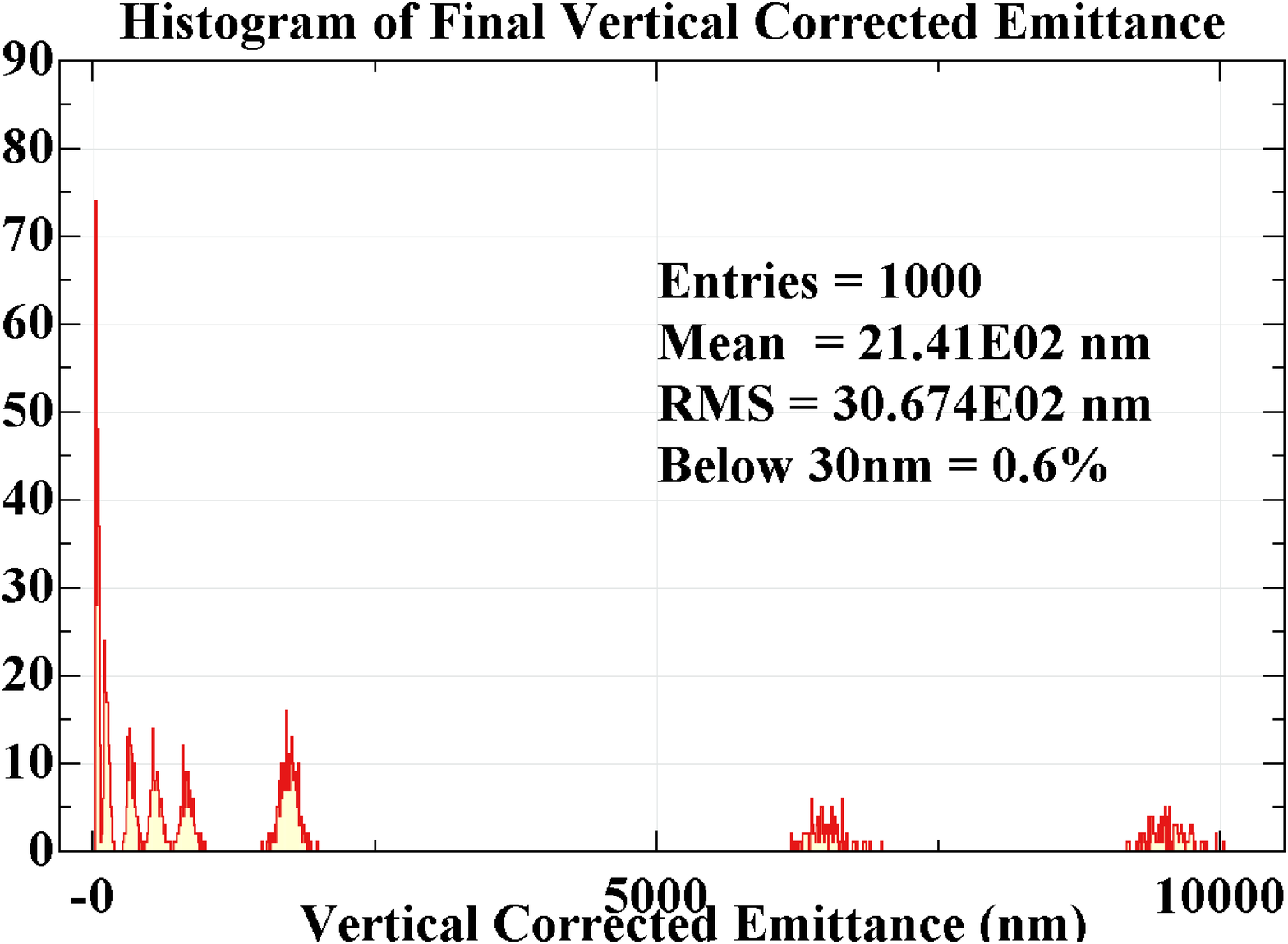}
	\caption{DMS results without PRM}
	\label{DMS_NoPRM}
\end{minipage}
\hspace{0.5cm}
\begin{minipage}[b]{0.5\linewidth}
	\centering
		\includegraphics[width=\columnwidth]{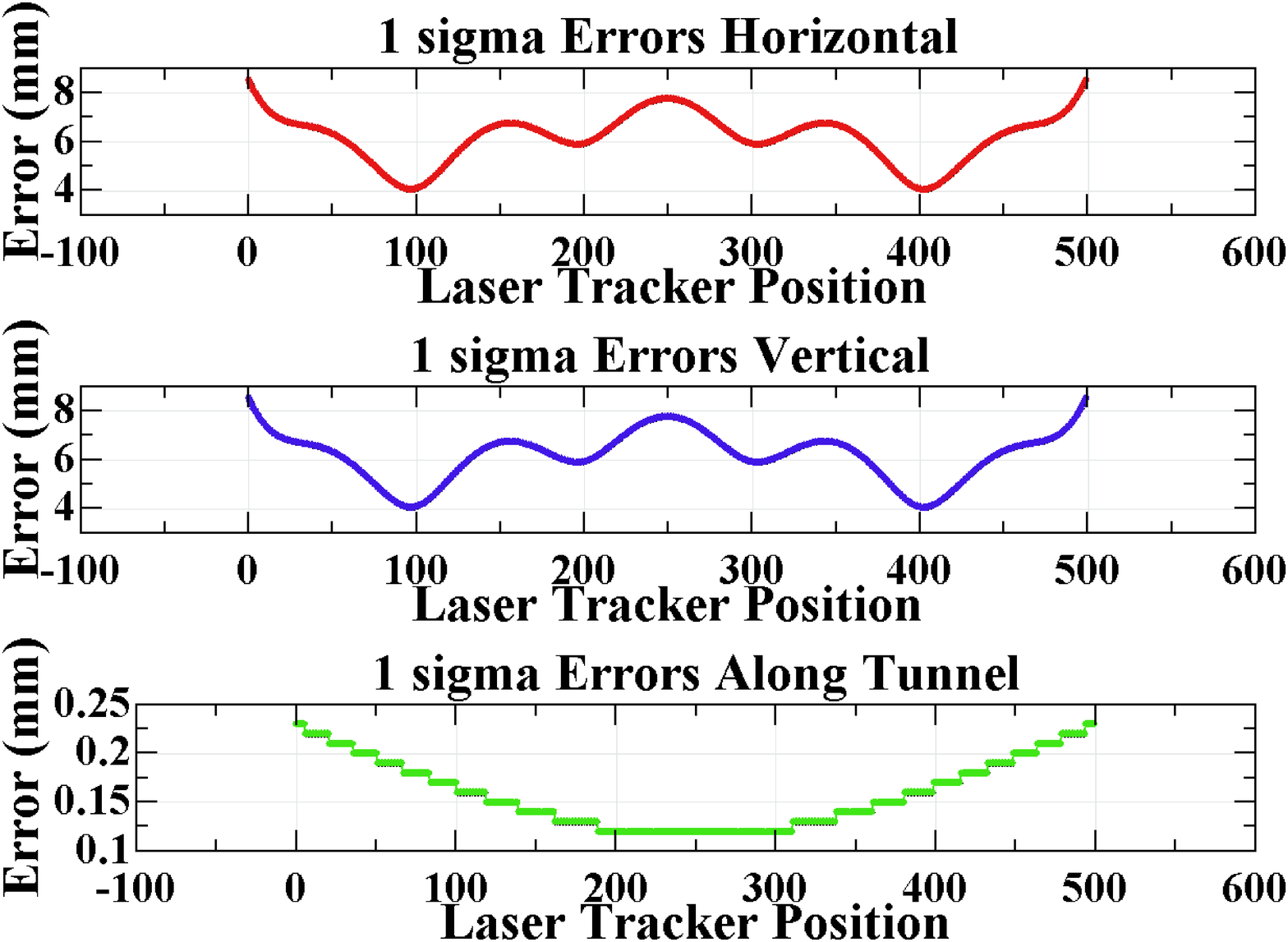}%
	\caption{The Errors on the reference network with PRM}
	\label{NetorkErrors_PRM}
	\end{minipage}
\end{figure}

For each of the ten simulated networks, 100 DMS simulations were performed with different random seeds.  The final vertical corrected emittance for each run is histogramed and shown in figure \ref{DMS_NoPRM}.

It is clear from figure~\ref{DMS_NoPRM} that a survey of the ILC with only laser trackers would not be sufficient as only 0.6\% of the simulations give a final corrected vertical emittance within the design specification for the ILC main linac (less than 30nm).  Figure~\ref{DMS_NoPRM} also shows that each simulated network results in a distinct group of final corrected vertical emittances.

\subsection{Simulations with PRM}

The results of simulations including PRM measurements, in the form of baseline vector differences in the network simulations, are discussed in sections \ref{adjustPRM} and \ref{DMSGPS}.

\subsubsection{Network Adjustment Results}
\label{adjustPRM}
As before PANDA was used to determine the errors on the adjusted network (see figure~\ref{NetorkErrors_PRM}).  The error curves shown in figure~\ref{NetorkErrors_PRM} follow the expected double minima curve, with a pull down effect caused by the PRMs. They also show the same similarity between horizontal and vertical errors.

%\begin{figure}
%\begin{minipage}[b]{0.5\linewidth}
%	\centering
%		\includegraphics[width=\columnwidth]{plots/PRMAdjustmentError.eps}%
%	\caption{The Errors on the reference network with PRM}
%	\label{NetorkErrors_PRM}
%	\end{minipage}
%\end{figure}

\begin{figure}
	\begin{minipage}[b]{0.5\linewidth} % A minipage that covers half the page
		\centering
		\includegraphics[width=\columnwidth]{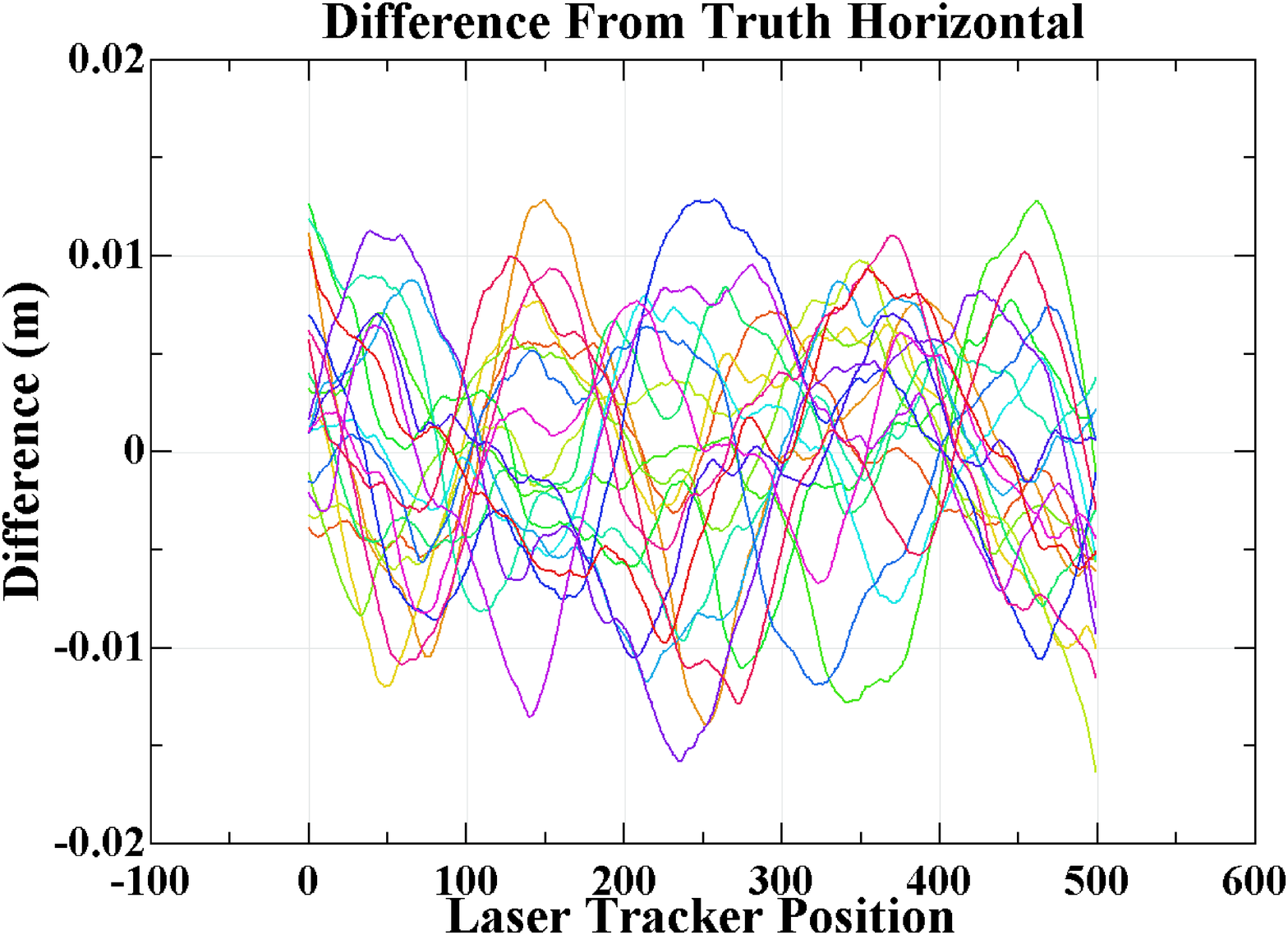}
		\caption{The Horizontal Difference from truth with PRM}
		\label{Horizontal_PRM}
	\end{minipage}
\hspace{0.5cm} % To get a little bit of space between the figures
	\begin{minipage}[b]{0.5\linewidth}
		\centering
		\includegraphics[width=\columnwidth]{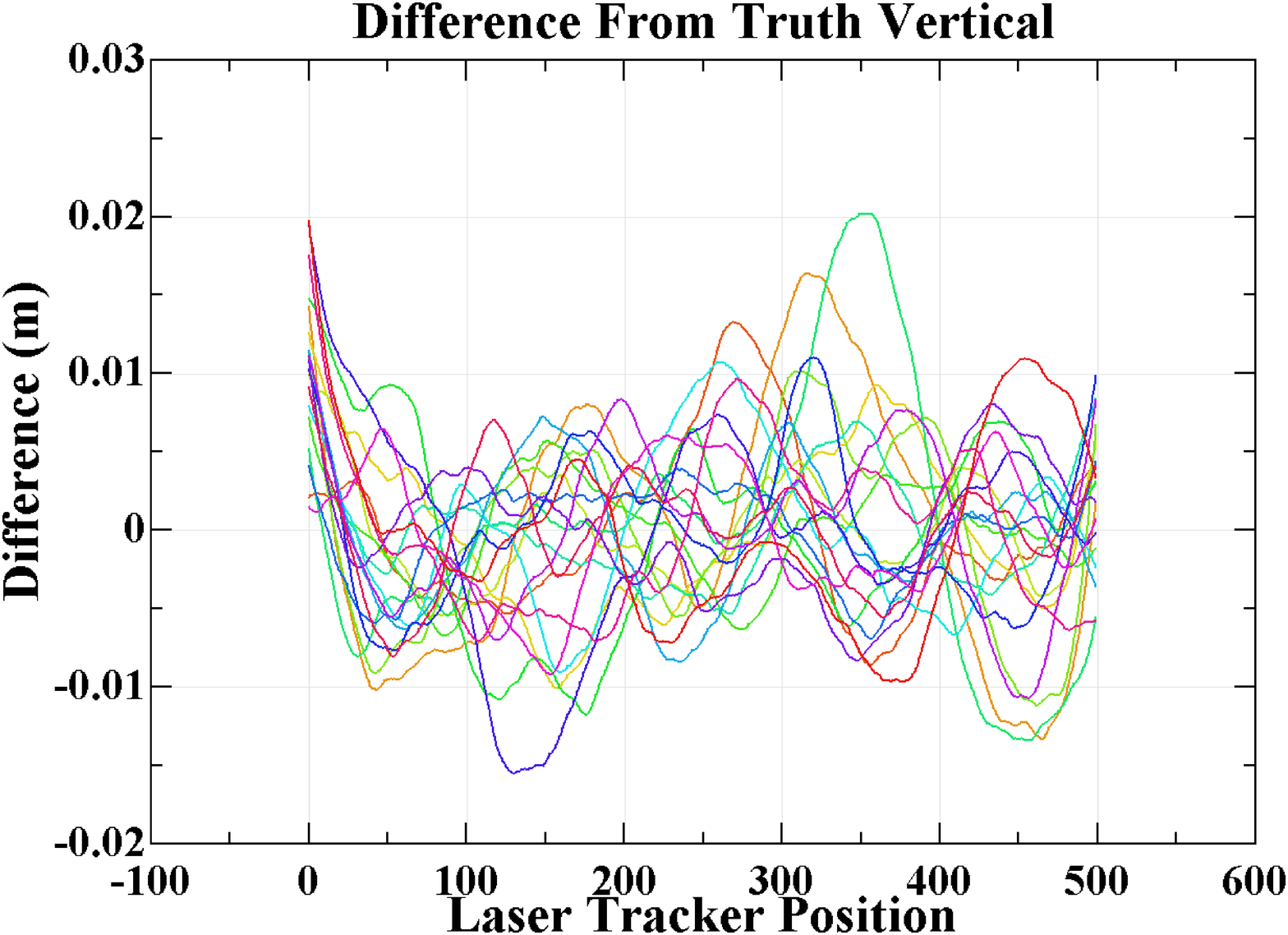}
		\caption{The Vertical Difference from truth with PRM}
		\label{Vertical_PRM}
	\end{minipage}
\end{figure}

Twenty reference network simulations were performed.  The truth was subtracted from the horizontal and vertical reconstructed positions and the difference plotted (see figures \ref{Horizontal_PRM} and \ref{Vertical_PRM} respectively).

The horizontal difference, shown in figure~\ref{Horizontal_PRM}, appears as expected given the horizontal errors shown in figure~\ref{NetorkErrors_PRM}; however the vertical difference from the truth, shown in figure~\ref{Vertical_PRM}, again does not appear as expected. Figure~\ref{Vertical_PRM} has a systematic curvature at the start indicating the same problem with the PANDA software seen in section \ref{adjustNoPRM}. Although the systematic curvature is a problem, the horizontal network can again be used to misalign the vertical plane of the simulated ILC.

\subsubsection{DMS Results}
\label{DMSGPS}

\begin{figure}
	\centering
		\includegraphics[width=0.5\columnwidth]{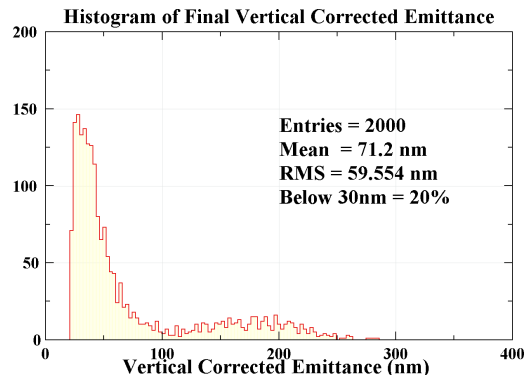}
	\caption{DMS results with PRM}
	\label{DMS_PRM}
\end{figure}

The process described in section \ref{dmsNoGPS} was repeated for the twenty simulated network adjustments with PRMs and the results shown in figure~\ref{DMS_PRM}.

%As before, for each of the twenty simulated network adjustments, 100 DMS simulations were performed. The horizontal adjustment results were used to misalign the vertical plane of the simulated accelerator.  The final vertical corrected emittance for each run is histogamed as shown in figure~\ref{DMS_PRM}.

The histogram (figure~\ref{DMS_PRM}) shows that the introduction of PRMs has caused a large improvement to the final corrected vertical emittance; however, only 20\% of the simulations are below the required 30nm.

\section{Rapid Tunnel Reference Survey Model (RTRSM)}
\label{RTRSM}

Full simulations of the ILC reference network require large linear algebra models which can take a long time to set up and solve. For beam dynamics simulations a large number of reference networks are required; so an attempt has previously been made $^{\cite{RandomWalk}}$ to create a model which can generate a reference network efficiently. The model should produce simulations of an ILC reference network rapidly and with the same statistical properties that a full simulation would have. The model is in essence a random walk model, but modified to take account of PRMs and systematic errors.  The errors of such fast simulations are examined in the following sections.

\subsection{RTRSM Error curves without PRM}

\begin{figure}
\begin{minipage}[b]{0.5\linewidth}
	\centering
		\includegraphics[width=0.9\columnwidth]{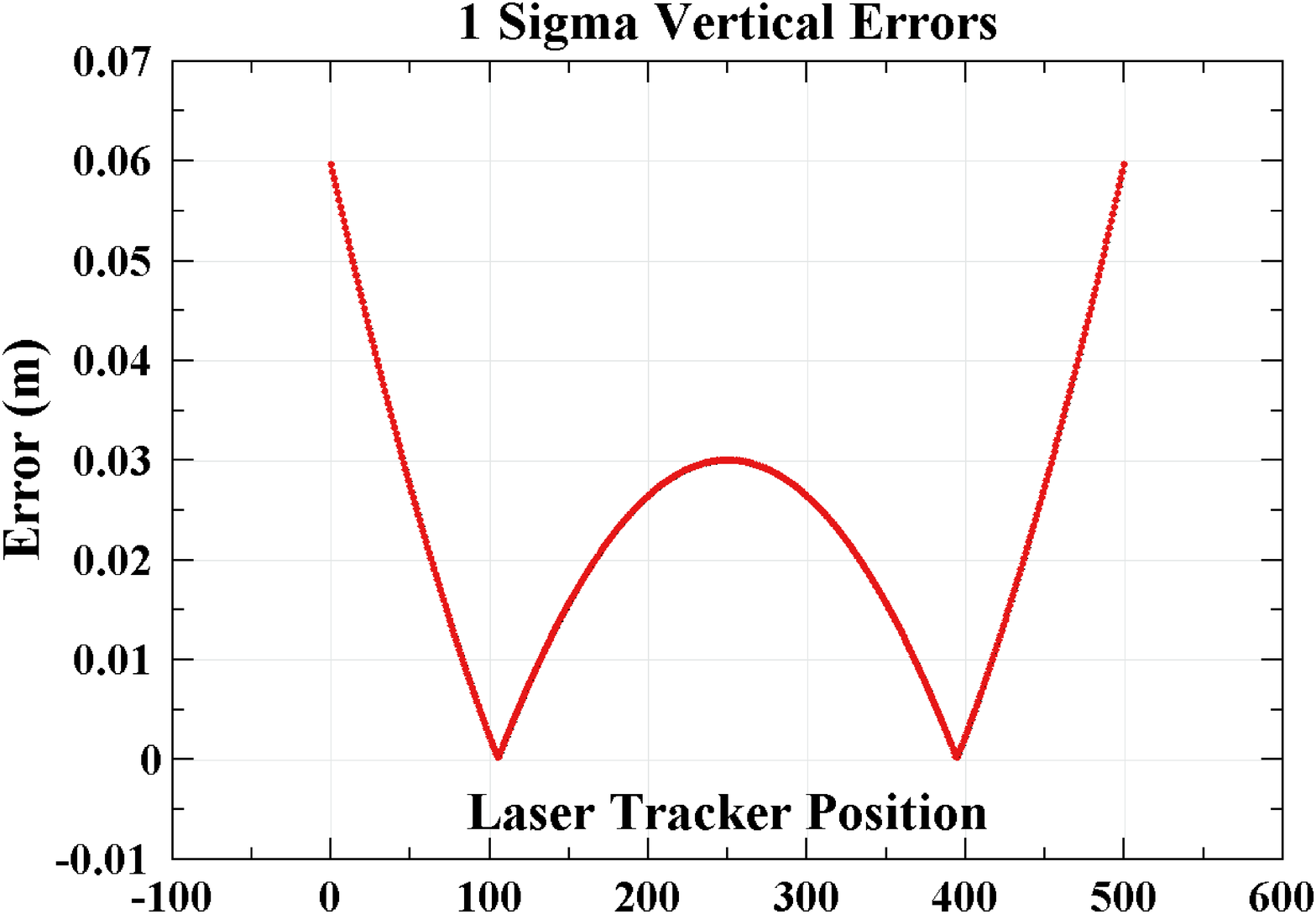}
	\caption{Reference network simulations model errors without PRM}
	\label{model_noPRM}
	\end{minipage}
	\hspace{0.5cm} % To get a little bit of space between the figures
	\begin{minipage}[b]{0.5\linewidth}
	\centering
		\includegraphics[width=0.9\columnwidth]{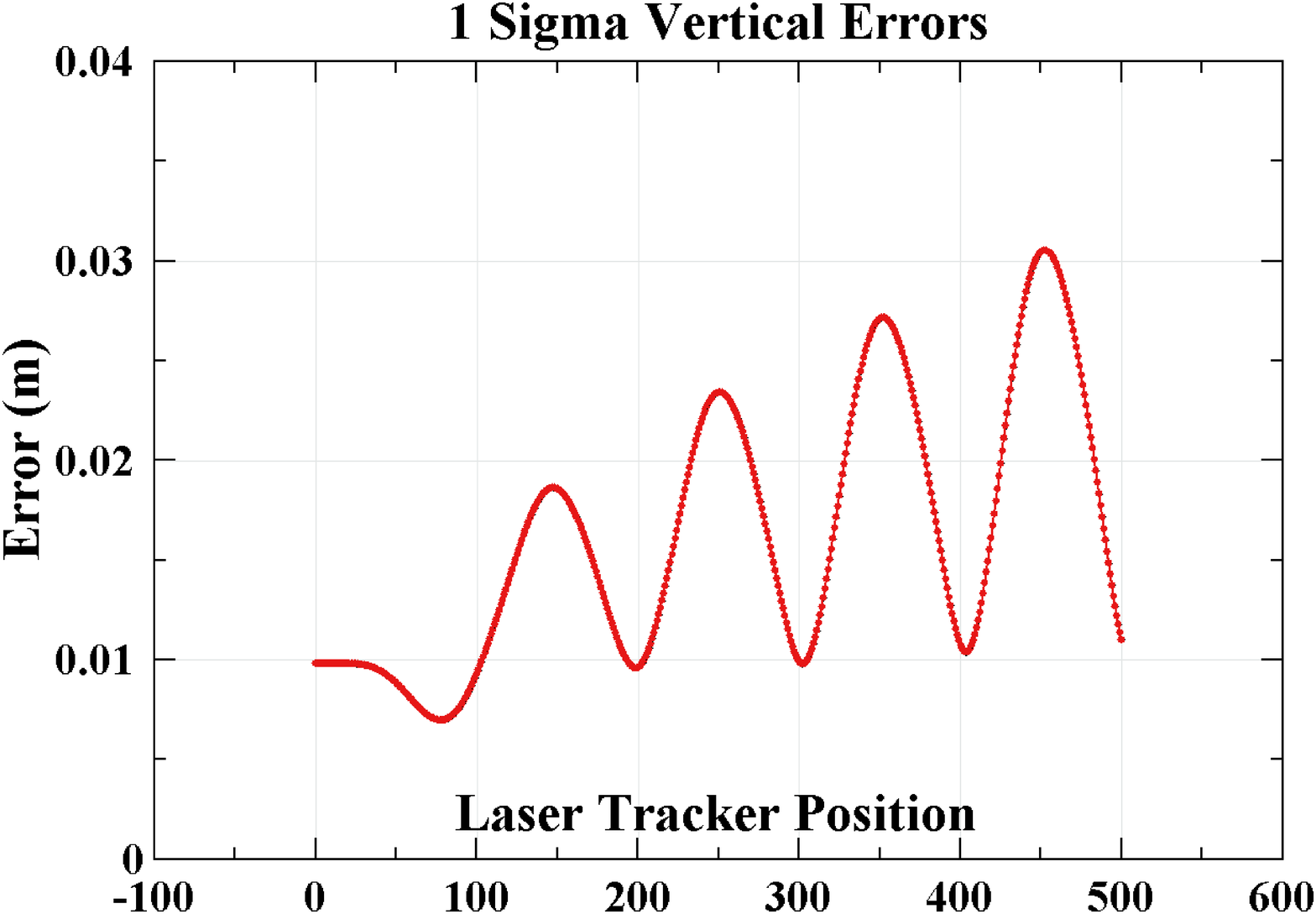}
	\caption{Reference network simulations model errors with PRM}
	\label{model_PRM}
	\end{minipage}
\end{figure}

The RTRSM is first studied without PRMs (see table \ref{tab:LETModelparameters} for parameters).  To understand the errors 2000 simulations using the RTRSM were performed with the systematic parameters varied by the amount defined in table \ref{tab:LETModelparameters}.  A straight line was fitted to each of the simulations and the residuals found; the residuals were then averaged at each point, and the standard deviations found and plotted in figure~\ref{model_noPRM}. Figure~\ref{model_noPRM} shows the standard double minima curve expected, only much sharper than the error curves in figure~\ref{NetorkErrors_NoPRM} due to the cominance of systematic errors.  In this mode the RTRSM works well.

\begin{table}
	\centering
		\begin{tabular}{|c |c |c |c|c|}
		\hline
			$a_{y}$ & $a_{\theta}$ & $\delta \theta_{syst}$ & $\delta y_{syst}$ & $l_{l}$\\
			\hline
			5 $\mu$m & 55.4 nrad & -260 nrad & -5.3 $\mu$m & 25 m\\
			
		\hline	
		\end{tabular}
	\caption{RTRSM parameters}
	\label{tab:LETModelparameters}
\end{table}

\subsection{RTRSM Error curves with PRM}

Two thousand simulations were performed using the RTRSM with PRMs added. As before a straight line was fitted to each simulation and the standard deviation found and shown in figure~\ref{model_PRM}.  The errors shown in figure~\ref{model_PRM} indicate a problem with the model; we would expect an error curve which resembles the one shown in figure~\ref{NetorkErrors_PRM}; a double minima pulled down by the PRMs. The error curve in figure~\ref{model_PRM} is asymmetric and of larger than expected magnitude.

The problem with the RTRSM most likely arises from the incorrect combination of errors.  The random walk starts with zero error which increases along the network, and the PRMs have errors spread evenly around a straight line.  The model attempts to combine both sets of data but does not take account of the frame the errors are in. This leads to the observed asymmetric behavior and larger than expected errors. The RTRSM has been used to do DMS simulations in \cite{Freddy_simul}; the findings here invalidate these results. Work has already commenced on an improved model.

\section{Conclusion}

Conventional alignment using laser trackers without GPS based primary reference markers, ignoring systematic errors, is not suitable for the ILC as only 0.6\% of all DMS simulations achieve the required emittance of less than 30nm. The introduction of primary reference markers does improve the alignment, however the results are still not good enough as only 20\% of simulated linacs fall below the required emittance.

From the study of the errors of the RTRSM, it can be seen that without PRMs the error curves are as expected, whereas the introduction of PRMs causes the error curves to behave in an incorrect way revealing a flaw in the RTRSM.  Therefore, the current model needs further development, which has already begun.

\section{Acknowledgements}
Markus Schlösser From DESY, Geodesy group for help with PANDA\\Freddy Poirier for his help in using MERLIN

\begin{footnotesize}
% IF YOU DO NOT USE BIBTEX, USE THE FOLLOWING SAMPLE SCHEME FOR THE REFERENCES
% ----------------------------------------------------------------------------

\end{footnotesize}

% ****************************************************************************
% END OF BIBLIOGRAPHY AREA
% ****************************************************************************

\end{document}